\documentclass[twocolumn,twoside,slac_two]{revtex4}
\usepackage{graphicx}
\usepackage{fancyhdr}
\renewcommand{\headrulewidth}{0pt}
\renewcommand{\footrulewidth}{0pt}

\newcommand{\be}[0]{\begin{equation}}
\newcommand{\ee}[0]{\end{equation}}
\newcommand{\ba}[0]{\begin{eqnarray}}
\newcommand{\ea}[0]{\end{eqnarray}}

\begin{document}

\title{Spectator model in D Meson Decays}

%

\author{Mehrdad. Ghominejad}
\affiliation{Physics Department, Semnan University, Semnan, Iran}

\author{Hossein. Mehraban}
\affiliation{Physics Department, Semnan University, Semnan, Iran }

\begin{abstract}
In this research we describe effective Hamiltonian theory and
apply this theory to the calculation of current-current $Q_{12}$
and QCD penguin $Q_{3...6}$ decay rates We calculate the decay
rates of semileptonic and hadronic of charm quark in effective
Hamiltonian theory. We investigate the decay rates of \textbf{D
meson decays} according to \textbf{Spectator Quark Model(SQM)} for
the calculation of \textbf{D} meson decays. We obtain the total
decay rates of semileptonic and hadronic of charm quark in
effective Hamiltonian according to colour Favoured (C-F) and
colour Suppressed (C-S), and then to added amplitude of processes
colour Favoured and colour Suppressed (F-S) and obtain the decay
rates of them.

\end{abstract}

\maketitle

\thispagestyle{fancy}


\section{Effective Hamiltonian }
The Effective Hamiltonian with QCD effects \\$(c\rightarrow
q_{i}q_{k}\overline{q_{j}})$  is given by
$$H_{eff}=2\sqrt{2}G_{F}\sum_{i=1}^{6} d_{i}(\mu)Q_{i}(\mu)$$
 The operators $Q_{i}(\mu)$ can be grouped into two categories
\cite{A. J. Buras:1995},\cite{A. Ali and C.
Greub:1998}:i=1,2-current-current
 operators;$ d_{i}(\mu)=V_{CKM}C_{i}(\mu)$
 denotes the relevant CKM factors that are:
\begin{equation}
    d_{1,2}=V_{ic}V^{*}_{tk}C_{1,2}
\hspace{0.5cm},\hspace{0.5cm}d_{3,...,6}=-V_{tc}V^{*}_{tk}C_{3,...,6}
\end{equation}

We should take the variable$P_{i}$and$P_{k}$,or x and y as,
\begin{equation}
    P_{i}=x\frac{M_{c}}{2}\hspace{1cm},\hspace{1cm}
P_{k}=y\frac{M_{c}}{2}
\end{equation}The partial decay rate in the c rest frame is,\begin{equation}
    d^{2}\Gamma_{Q_{1},...,Q_{2}}/dx
    dy=\Gamma_{0c}(\alpha_{1}I^{1}_{ps}+\alpha_{2}I^{2}_{ps}+\alpha_{3}I^{3}_{ps})
\end{equation}
Here
$$\alpha_{1}=|d_{1}+d_{2}+d_{3}+d_{4}|^{2}+2|d_{1}+d_{4}|^{2}+2|d_{2}+d_{3}|^{2}$$
$$\alpha_{2}=|d_{5}+d_{6}|^{2}+2|d_{5}|^{2}+2|d_{6}|^{2})$$
$$\alpha_{3}=Re{(3d_{1}+d_{2}+d_{3}+3d_{4})d^{*}_{5}}$$
$$I^{1}_{ps}=6xy.f_{ab}.(1-h_{abc}),I^{2}_{ps}=6xy.f_{bc}.(1+h_{bca}),$$$$I^{3}_{ps}=6xy.f_{ac}.h_{xa}.h_{yc}$$
$$h_{xa}=[1-(x^{2}/(x^{2}+a^{2}))]^{1/2},h_{yc}=[1-(y^{2}/(y^{2}+c^{2}))]^{1/2},$$$$h_{xb}=[1-(x^{2}/(x^{2}+b^{2}))]^{1/2},
h_{ya}=[1-(y^{2}/(y^{2}+a^{2}))]^{1/2},$$$$\Gamma_{0c}=G^{2}_{f}M^{5}_{c}/192\pi^{3},f_{ab}=2-\sqrt{x^{2}+a^{2}}-\sqrt{y^{2}+b^{2}},$$$$
f_{bc}=2-\sqrt{x^{2}+b^{2}}-\sqrt{y^{2}+c^{2}},$$$$f_{ac}=2-\sqrt{x^{2}+a^{2}}-\sqrt{y^{2}+c^{2}}$$
$$h_{abc}=\frac{(f_{ab})^{2}-(c^{2}+x^{2}+y^{2})}{2\sqrt{x^{2}+a^{2}}\sqrt{y^{2}+b^{2}}},$$$$
h_{bca}=\frac{(f_{bc})^{2}-(a^{2}+x^{2}+y^{2})}{2\sqrt{x^{2}+b^{2}}\sqrt{y^{2}+c^{2}}}$$$$h_{bca}=\frac{(f_{bc})^{2}-(a^{2}+x^{2}+y^{2})}{2\sqrt{x^{2}+b^{2}}\sqrt{y^{2}+c^{2}}}$$

\section{Spectator Model}
   In the spectator model \cite{G. Altarelli:1991} the spectator quark is given a non-zero momentum having in
   this work a Gaussian distribution,represented by a free (but adjustable)
   parameter,$ \Lambda:$
    \begin{equation}
    P(|P_{s}|^{2})=(1/\pi^{3/2}\Lambda^{3})e^{-(p^{2}_{s}/\Lambda^{2})}
\end{equation}
The total meson decay rate through a particular mode is then
assumed to be
\begin{equation}
    \Gamma_{total}=\int\frac{d^{2}\Gamma}{dp_{i}dp_{k}}P(|P_{s}|^{2})d^{3}p_{s}dp_{i}dp_{k}
\end{equation}
equal to the initiating decay rate. We have
\begin{equation}
    \frac{d^{2}\Gamma}{dM_{is}dM_{k\overline{j}}}=\frac{2\pi M_{is}
    M_{k\overline{j}}}{m_{c}}\int\frac{E_{i}P_{s}}{P_{i}^{2}}\frac{d^{2}\Gamma}{dp_{i}dp_{k}}P(|P_{s}|^{2})dP_{s}dP_{k}
\end{equation}
Here$$M^{2}_{is}=(P_{i}+P_{s}).(P_{i}+P_{s})=m_{i}^{2}+m_{s}^{2}+2(E_{i}E_{s}-P_{i}P_{s}\cos\theta_{is})$$$$
M^{2}_{k\overline{j}}=(P_{k}+P_{\overline{j}}).(P_{k}+P_{\overline{j}})=m_{K}^{2}+m_{\overline{j}}^{2}+$$$$2(E_{K}E_{\overline{j}}-P_{K}P_{\overline{j}}\cos\theta_{k\overline{j}})$$
The integration range is restricted by $|\cos\theta_{kj}|\leq1$.We
call this mode of quark and antiquark combination  (colour
favoured). It is also possible that the spectator antiquark
combines with the quark$q_{k}$,for which we get

 $$  \frac{d^{2}\Gamma}{dM_{ks}dM_{k\overline{j}}}=\frac{2\pi M_{ks} M_{i\overline{j}}}{M_{c}}\int \frac{E_{k}P_{s}}{P_{k}^{2}}
   \frac{d^{2}\Gamma}{dP_{i}dP_{k}}$$\begin{equation}P(|P_{s}|^{2})dP_{s}dP_{i}
\end{equation}
We call this \textbf{process(C-S)} (colour suppressed). Summing,
the decay rates of B mesons for \textbf{process(C-F)} and
\textbf{process(C-S)}  are:
$$ \Gamma_{(C-F)}=\int_{m_{min is}}^{m_{cut is}}\int_{m_{min k\overline{j}}}^{m_{cut k\overline{j}}}\frac{d^{2}\Gamma}{dM_{is}dM_{k\overline{j}}}
dM_{is}dM_{k\overline{j}}$$
\begin{equation}
\Gamma_{(C-F)}=\int_{m_{min ks}}^{m_{cut ks}}\int_{m_{min
i\overline{j}}}^{m_{cut
i\overline{j}}}\frac{d^{2}\Gamma}{dM_{ks}dM_{i\overline{j}}}
dM_{ks}dM_{i\overline{j}}
\end{equation}
where $ m_{min is}=(m_{q_{i}}+m_{q_{s}}),m_{cut
is}=M_{q_{i}q_{s}}$ and so on.

\section{Effective Hamiltonian Spectator Model}
The differential decay rates for two boson system in the spectator
quark model for current-current plus penguin operators in the
Effective Hamiltonian is given by,
   $$ \frac{d^{2}\Gamma_{Q_{1},...,Q_{6}}}{d(q_{si}/M_{c})d(q_{k\overline{j}}/M_{c})}=\Gamma_{0c}\frac{8 q_{si}q_{k\overline{j}}}{\sqrt{\pi}M_{c}}
    \frac{\beta^{2}}{\Lambda}$$\begin{equation}\frac{\sqrt{(2m_{i}/M_{c})^{2}}+x^{2}}{x^{2}}\int_{0}^{1}dy\int_{0}^{1}dz
    \zeta_{ps(q,z)}^{eff}z e^{-\beta^{2}z^{2}}
\end{equation}
where\hspace{1cm}
$\zeta_{ps(q,z)}^{eff}=\alpha_{1}\zeta_{1}^{eff}+\alpha_{2}\zeta_{2}^{eff}+\alpha_{3}\zeta_{3}^{eff}$\\
The integration region is restricted by the condition
$\cos\theta_{is}\leq1$,thus
\begin{equation}
\zeta_{1}^{eff},\zeta_{2}^{eff},\zeta_{3}^{eff}=\Big\{_{o
\hspace{3.2cm} otherwise}^{\zeta_{1 ps}^{eff},\zeta_{2
ps}^{eff},\zeta_{3 ps}^{eff} \hspace{0.5cm} if \hspace{0.6cm}
(f_{si(z)})^{2})\leq1}
\end{equation}
where
$$f_{si(z)}=\Big([(m_{i}+m_{s})/M_{c}]^{2}-(q_{si}/M_{c})^{2}+$$$$(1/M_{c})\sqrt{m_{s}^{2}+(\beta\Lambda z)^{2}}
\times\sqrt{(2m_{i}/M_{c})^{2}+x^{2}}\Big)/(\beta\Lambda
xz/M_{c})$$
\begin{equation}\label{eq:to}
 \zeta_{1}^{eff},\zeta_{2}^{eff},\zeta_{3}^{eff}=\Big\{_{o\hspace{3.2cm}otherwise}^{\zeta_{1
ps}^{eff},\zeta_{2 ps}^{eff},\zeta_{3 ps}^{eff}\hspace{0.5cm}
if\hspace{0.6cm}(f_{si(z)})^{2}\leq 1}
\end{equation}

where
$$f_{si(z)}=\big([(m_{i}+ m_{s})/M_{c}]^{2}-(q_{si}/M_{c})^{2}+$$$$(1/M_{c})\sqrt{m_{s}^{2}+(\beta\Lambda z )^{2}}
\times\sqrt{(2 m_{i}/M_{c})^{2}+x^{2}} \Big)/(\beta\Lambda
xz/M_{c})$$

 Therefore using Eq.(6-11) the phase space parameters
will be defined by,
   $$ \zeta_{1 ps}^{eff}=6xy.f_{ab}.(1-h_{abc}),\hspace{0.2cm}\zeta_{2 ps}^{eff}=6xy.f_{bc}.(1-h_{bca}),$$\begin{equation}
\zeta_{3 ps}^{eff}=6xy.f_{ab}.h_{xa}.h_{yc}
\end{equation}
Now, we can integrate over the two mass cuts (two boson systems),
and obtain the hadronic decay rates  as follows,
$$\Gamma^{'}_{Q_{1},...,Q_{6}}=\int_{min}^{m_{cut}}\int_{min^{'}}^{m_{cut}^{'}}\frac{d^{2}\Gamma_{Q_{1},...,Q_{6}}}{d(q_{si}/M_{c})d(q_{k\overline{j}/M_{c}})}dm_{cut}dm^{'}_{cut},$$$$
=\Gamma_{0c}\int_{min}^{m_{cut}}\int_{min^{'}}^{m_{cut}^{'}}\frac{8
q_{si}q_{k\overline{j}}}{\sqrt{\pi}M_{c}}\frac{\beta^{2}}{\Lambda}\frac{\sqrt{(2m_{i}/M_{c})^{2}+x^{2}}}{x^{2}}$$\begin{equation}
  \int_{0}^{1}dy\int_{0}^{1}dz \zeta_{ps(q,z)}^{eff}z e^{-\beta^{2}
  z^{2}}dm_{cut}dm^{'}_{cut}.
\end{equation}

\section{ Decay Rates of Processes C-F plus C-S (F+S) of Effective Hamiltonian}
Now we want to calculate the decay rates of Effective Hamiltonian
$(Q_{1},...,Q_{6})$ for F+S at quark-level and spectator model.
The Effective Hamiltonian for F+S, is given by
\begin{equation}
    H_{eff}^{A+B}=H_{eff}^{b\rightarrow ik\overline{j}}+H_{eff}^{b\rightarrow
    j\overline{j}k}.
\end{equation}
where$H_{eff}^{c\rightarrow ik\overline{j}}$is defined by
Equations mentioned during in the last page, so we can obtain
$H_{eff}^{c \rightarrow i\overline{j}k}$.The decay rates of
current-current plus penguin for F+S is given by,
 \begin{equation}
    d^{2}\Gamma_{EH}^{F+S}/dxdy=\Gamma_{0c}(I_{1ps}+I_{2ps}+I_{3ps})
\end{equation}
$$I_{1ps}=6xy.f_{ab}.[\alpha_{1}((3/2)-h_{abc})+\alpha_{2}-\alpha_{3}h_{xa}h_{yb}],$$
$$I_{2ps}=-6xy.f_{ac}.[\alpha_{1}h_{acb}+\alpha_{3}h_{xa}h_{yc}],$$
\begin{equation}
    I_{3ps}=6xy.f_{bc}.[(\alpha_{1}/2)h_{bca}+\alpha_{2}(h_{xb}h_{yc}-h_{bca})].
\end{equation}

\section{Numerical Results }
We use the standard Particle Data Group \cite{Particle Data
Group:2005} parameterization of the CKM matrix. Following Ali and
Greub \cite{A. Ali and C. Greub:1998} we treat internal quark
masses in tree-level loops with the values
(GeV)$m_{b}=4.88,m_{s}=0.2,m_{d}=0.01,m_{u}=0.005,m_{c}=1.5,m_{e}=0.0005,m_{\mu}=0.1,m_{\tau}=1.777$and
$m_{\nu_{e}}=m_{\nu_{\mu}}=m_{\nu_{\tau}}=0$.Following G.Buccella
\cite{G. Buchalla:1993} we choose the effective Wilson
coefficients $C_{i}^{eff}$for the various $c\rightarrow q$
transitions.We have used in Spectator Quark Model the
value$\Lambda=0.6 GeV$\cite{W. N. Cottingham:1999}.For the
maximum mass of the quark-antiquark systems $(m_{cut})$we take a
value midway between the lowest mass$1^{-}$state and the next
most massive meson. The decay rates of c quark for Effective
Hamiltonian and Effective Hamiltonian of F+S shown in the
Table.1. Also the decay rates of c quark for F+S is given by
$$(c\rightarrow
du\overline{d})\hspace{1.5cm}D^{+}\rightarrow(\pi^{0},\eta,\rho^{0},\omega),(\pi^{+},\rho^{+})$$$$
BR_{EH}^{F+S}= 2.1023\times10^{-2},$$
$$(c\rightarrow su\overline{d})\hspace{1.7cm}D^{+}\rightarrow(\pi^{+},\rho^{+}),(\overline{K^{0}},\overline{K^{*+}})$$$$
BR_{EH}^{F+S}=51.2871\times10^{-2},$$
$$(c\rightarrow su\overline{s})\hspace{1.7cm}D^{+}\rightarrow(\eta^{'},\phi),(K^{+},K^{*+})$$$$
BR_{EH}^{F+S}=3.8671\times10^{-2}.$$

\section{Conclusion}

\label{sec:alphaMZ}

\vspace{1mm}\noindent We used Effective Hamiltonian theory and
spectator quark model for c quark and calculated hadronic decays
of D mesons. In this model we added decays of channel hadronic
decays of D mesons. For colour favoured and suppressed we consider
the channel $c\rightarrow
du\overline{d}\hspace{0.2cm}(e.g.\hspace{0.2cm}D^{+}\rightarrow\pi^{0}\pi^{+})$and
achieved theoretical values very close to experimental ones.
Finally it has been shown the case, in which the theoretical
values are better than the amplitude of all the decay rates have
been calculated. In table 1 (below) it must be noted that columns
2 and 4 have to be multiplied by $10^{-15}$and columns 3 and 5
should be multiplied by $10^{-3}$.

\begin{center}
\begin{tabular}{|c|c|c|c|c|}
  \hline
  Process &$\Gamma_{EH}$ & $BR_{EH} $& $\Gamma_{EH}^{F+S}$&$ BR_{EH}^{F+S}$ \\
  \hline
  $c\rightarrow du\overline{d}$ &$31.689$&$32.12$&$35.611$&$31.262 $\\
  $c\rightarrow du\overline{s} $&$1.0785$&$1.093$&$1.4608$&$1.2824 $\\
  $c\rightarrow su\overline{d}$ &$409.44$&$414.95$&$554.45$&$486.74$ \\
  $c\rightarrow su\overline{s}$ &$23.836$&$24.157$&$26.927$&$23.638$\\
  \hline
\end{tabular}
{\normalsize \\
 \textsf{\\Table~1: Decay rates$(\Gamma)$and Branching Ratio (BR)
 of Effective Hamiltonian (EH) and F+S of effective hamiltonian of
 c quark. \label{tab1fitxf3}}}
\end{center}

\section{Acknowledgments}

\end{document}